\documentclass{jfm}

\usepackage{graphicx}
\usepackage{natbib}

\ifCUPmtlplainloaded \else
  \checkfont{eurm10}
  \iffontfound
    \IfFileExists{upmath.sty}
      {\typeout{^^JFound AMS Euler Roman fonts on the system,
                   using the 'upmath' package.^^J}%
       \usepackage{upmath}}
      {\typeout{^^JFound AMS Euler Roman fonts on the system, but you
                   dont seem to have the}%
       \typeout{'upmath' package installed. JFM.cls can take advantage
                 of these fonts,^^Jif you use 'upmath' package.^^J}%
      }
  \else
  \fi
\fi

\ifCUPmtlplainloaded \else
  \checkfont{msam10}
  \iffontfound
    \IfFileExists{amssymb.sty}
      {\typeout{^^JFound AMS Symbol fonts on the system, using the
                'amssymb' package.^^J}%
       \usepackage{amssymb}%
         
       \let\ge=\geqslant  
      }{}
  \fi
\fi

\ifCUPmtlplainloaded \else
  \IfFileExists{amsbsy.sty}
    {\typeout{^^JFound the 'amsbsy' package on the system, using it.^^J}%
     \usepackage{amsbsy}}
    {}
\fi

\newcommand\St{\mbox{St}}
\newcommand\principal{\mbox{p.v.}}

\newcommand\Ai{\mbox{Ai}}            
\newcommand\Bi{\mbox{Bi}}            

\newsavebox{\astrutbox}
\sbox{\astrutbox}{\rule[-5pt]{0pt}{20pt}}

\title[Phase transitions in the distribution of inelastically colliding inertial particles]{Phase transitions in the distribution of inelastically colliding inertial particles}

\author[S. Belan, A. Chernykh and G. Falkovich]%
{S. Belan$^{1,2}$\thanks{Email address for correspondence: belan@itp.ac.ru}, \ns
A. Chernykh$^{3,4}$\break
and G. Falkovich$^{5,6}$
}

\affiliation{$^1$ Moscow Institute of Physics and Technology, Dolgoprudny, Russia\\[\affilskip]
$^2$Landau Institute for Theoretical Physics, Chernogolovka, Russia\\[\affilskip]
$^3$Institute of Automation and Electrometry, Novosibirsk, Russia\\[\affilskip]
$^4$Novosibirsk State University, Novosibirsk, Russia\\[\affilskip]
$^5$Weizmann Institute of Science, Rehovot,  Israel\\[\affilskip]
$^6$Institute for Information Transmission Problems, Moscow, Russia}

\pubyear{2010}
\volume{650}
\pagerange{119--126}
\date{?; revised ?; accepted ?. - To be entered by editorial office}
\begin{document}

\maketitle

\begin{abstract}
It was recently suggested that the sign of particle drift in inhomogeneous temperature or turbulence depends on the particle inertia: weakly inertial particles localize near minima of temperature or turbulence intensity (effects known as thermophoresis and turbophoresis), while strongly inertial particles fly away from minima in an unbounded space. The problem of a particle near minima of turbulence intensity is related to that of two particles in a random flow, so that the localization-delocalization transition in the former corresponds to the path-coalescence transition
in the latter.
 The transition is signaled by the sign change of the Lyapunov exponent that characterizes the mean rate of particle approach to the minimum (which could be a wall or another particle). Here we solve analytically this problem for inelastic collisions and derive the phase diagram for the transition in the inertia-inelasticity plane.
 An important feature of the phase diagram is the region of inelastic collapse: if the restitution coefficient of particle velocity is smaller than some critical value, then the particle is localized for any inertia.
  We present direct numerical simulations which support the theory and in addition reveal the dependence of the transition of the flow correlation time, characterized by the Stokes number.  \end{abstract}

\section{Introduction}\label{sec:introduction}
Transport of inertial particles in an inhomogeneous turbulence or in a temperature gradient is of great importance for various industrial and natural processes.
There is the widely known thermophoresis: the tendency of the particles to migrate in the direction of decreasing temperature \citep{Kampen,Milligen}. A similar phenomenon for inertial particles in non-uniform turbulence is called turbophoresis \citep{Caporaloni_1975,Reeks_1983,Reeks_2014}.
The standard turbophoresis was consistently  described within the gradient transport model for the particle concentration under the assumption of local equilibrium \citep{Reeks_1983}.
In that model, the turbophoretic velocity arises as a specific term in the expression for particle current which is directly proportional to gradient of the local turbulence intensity.
The turbophoretic velocity is directed towards decreasing turbulence so that given zero concentration gradients and no net fluid flow in any direction, particles will migrate from high to low turbulence intensities.
If turbophoresis is strong enough to overcome the effect of turbulent diffusion, the particles turn out to be localized near the  minimum of turbulence.
It was recently suggested \citep{BFF} that particle migration in an unbounded space could be actually opposite- away from minima, if the particles are inertial enough.  
More specifically, the phenomenon of \emph{reverse turbophoresis}  takes place for the inertial particles placed in the vicinity of turbulence minimum provided that the particle mean free path is larger than distance from this minimum.
In this case the local equilibrium assumption fails and there is no reduced description of particle transport in terms of  spatial concentration. Thus, in contrast to the standard turbophoresis, the reverse turbophoresis is a nonlocal phenomenon, not associated with any local turbophoretic current.

The change of direction of particle migration leads to localization-delocalization transition upon the change of inertia. It also means separation: when time goes to infinity, particles with low inertia go to a minimum of turbulence and concentrate there, while sufficiently inertial particles escape to infinity.
In many cases of interest, there is a wall, which corresponds to the minimum either of turbulence (in wall-bounded flows) or of temperature (in furnaces, combustion chambers and kerosene lanterns). It is not known how the boundary conditions at the wall affect the direction of the particle drift. Here we derive an analytical expression for the Lyapunov exponents associated with the motion of inertial particles near an inelastic wall. That allows us to predict the localization-delocalization transition upon the change in elasticity of collisions. A central result of our work is a phenomenon which one might call \emph{inelastic collapse}: if the restitution coefficient of particle velocity is smaller than the critical value that we determine, then the particle is localized near the wall for any inertia.
The theoretical predictions are in a good agreement with the results of  numerics that we carry.

All the results are directly translated into the description of the statistics of the distance between two inertial particles in a spatially smooth and temporally random one-dimensional flow \citep[see e.g.][]{WM,Bec_2008}. The problem of relative dispersion  is of particular importance for the description of distribution and collisions of water droplets in clouds and for the description of planet formation.
Here localization means that particles tend to approach each other and create clusters \citep{WM}.
Our findings mean that in one-dimensional flow particles always create clusters when  the restitution coefficient is below a threshold.
That may be of importance for wide classes of phenomena in industry, geophysics and astrophysics, from clouds to planet formation.
Note, however, that it will require further work to establish quantitatively how significant are the effects of inelastic inter-particle collisions in higher-dimensional flows.

\section{General relations}
\label{sec:general relations}
Consider the motion of a heavy particle embedded in an incompressible fluid flow ${\bf u}({\bf r},t)$ near flat impenetrable wall.
We introduce a reference system with the $z$-axis perpendicular to the wall and assume that the fluid occupies the region $z > 0$.
The particle is assumed to be so small that the flow around it is viscous.
Then, coordinate ${\bf r}$ and velocity ${\bf v}$ of the particle change according to
\begin{equation}
\label{Langeven}
\frac{d{\bf r}(t)}{dt}={\bf v}(t),\ \ \ \ \ \  \frac{d{\bf v}(t)}{dt}=\frac{ {\bf u}({\bf r}(t), t)-{\bf v}(t)}{\tau}.
\end{equation}
where $\tau$ is the particle response time (Stokes time).
 The boundary condition at $z=0$ is inelastically reflecting so that at every collision the particle loses a definite part of its wall-normal velocity $v \to -\beta v$, where  $v\equiv v_z$ and $\beta$ is the constant coefficient of restitution.

The fluid velocity field is treated as a random function of time. Its statistics is assumed to be homogeneous (stationary) in time, whereas there is no homogeneity in space:  typical amplitude of velocity fluctuations is assumed to depend on $z$-coordinate due to the presence of wall.
The joint probability density function (PDF) of the particle's velocity and
coordinate along the direction of flow inhomogeneity is defined as
\begin{equation}
\label{PDF definition}
\rho(z, v, t)=\langle \delta(z-z(t))\delta(v-v(t))\rangle,
\end{equation}
where $z(t)$ and $v(t)$ are the particular solutions of (\ref{Langeven}) and the averaging is over the statistics of random flow.

There are three characteristic times for an inertial particle in spatially inhomogeneous random flow: the velocity correlation time $\tau_c$ of the fluid, the particle relaxation time $\tau$, and the time needed by the particle to feel the flow inhomogeneity $\tilde\tau$, specified below.
We first consider the case when $\tau_c\ll \tau,\tilde \tau$, i.e. in particular, the Stokes number $\St=\tau/\tau_c\gg1$.
Then the fluid velocity field can be treated as short-correlated and  PDF (\ref{PDF definition}) is the solution of the Fokker-Planck
\begin{equation}
\partial_t \rho = -v \partial_z \rho+\gamma\partial_{v }(v \rho)+\gamma^2\kappa(z)\partial_{v }^2\rho\ ,\label{FP1}
\end{equation}
where $\gamma=1/\tau$ and the effective diffusivity $\kappa(z)=\int_0^\infty\langle u_z(z,t)u_z(z,0)\rangle dt$ describes the non-uniform intensity of turbulence.
The boundary condition at the inelastic wall is dictated by  particle number (or probability) conservation: the outcoming flux of particles at the boundary is balanced by incoming flux.
Thus
\begin{equation}
\label{boundary condition}
 \rho(z=0,v,t)=\beta^{-2} \rho(z=0,-v/\beta,t) \quad \mathrm{for}\ v>0.
 \end{equation}

Our paper is devoted to the case of the quadratic profile of the diffusivity
\begin{equation}
\label{diffusivity}
\kappa(z)=\mu z^2,
\end{equation}
 where $\mu$ measures the intensity of fluid velocity fluctuations.
 This model serves as a generic profile of the eddy diffusivity near a minima of intensity of random flow. Note, however, that for large-Reynolds wall-bounded turbulence, such model can have only qualitative use, since the effective diffusivity behaves as $z^4$ in the viscous sub-layer and as $z$ in the logarithmic boundary layer \citep{MY}.


The equations (\ref{FP1}) and (\ref{diffusivity}) also constitute a standard one-dimensional model to describe the PDF for relative motion of two particles at  viscous scale of turbulence \citep[]{FFS,Bec_2008}. In this case, $z$ is the inter-particle distance and $v$ is the relative velocity, so that $dz/dt=v$, $\tau dv/dt=-v +\Delta u$. At small separation $z$ one obtains $\Delta u(z,t) =z s(t)$ where $s(t)$ is the $z$-independent  gradient of fluid  velocity.
 Assuming $s(t)$ to be short-correlated we find Eq. (\ref{FP1}) with $\kappa(z) =\mu z^2$ and $\mu=\int_{0}^{\infty} \langle s(0)s(t)\rangle dt$.
The boundary condition (\ref{boundary condition}) takes into account inelastic inter-particle collisions.
Thus, the results of further consideration are applicable also to the problem of relative dispersion of inelastically colliding  particles in one-dimensional random flow with zero time correlation.
Note that in higher dimensions the longitudinal (radial) relative motion of particles is coupled with the transversal dynamics \citep{Piterbarg_2002, Bec_2007,Zaichik_2008}. 
For this reason, the influence of inelastic collisions on evolution of inter-particle distance in higher-dimensional flows requires a separate investigation.

The time evolution of an arbitrary initial  phase-space distribution is determined by Eqs. (\ref{FP1}) and (\ref{boundary condition}).
However, an exact analytic solution of this non-stationary problem is known only for the trivial case of uniform diffusivity and ideally reflecting wall ($\beta=1$). 
In \citep{Devenish_1999} the stationary solution for constant $\kappa$ and arbitrary $\beta$ was considered under the assumption  of a homogeneous distribution of particles in the core of flow.  
The results of their steady-state analysis indicate that inelastic boundary collisions lead to particle concentration near the wall. 
In general case of non-uniform turbulence the space-dependent $\kappa$  determines a local time scale $\tilde\tau(z)$ given by the time it takes the particle flying with a rms velocity to experience a difference in average turbulence intensity.
The time $\tilde\tau(z)$ is estimated as the spatial scale $l\simeq\kappa/\kappa'$ 
 of flow inhomogeneity divided
by the mean local "thermal" particle velocity, which can be obtained by comparing the second and last terms in the rhs
of (\ref{FP1}), $\bar v(z)\simeq (\gamma \kappa)^{1/2}$. The validity of  (\ref{FP1}) requires that $\tau_c\ll\tilde\tau(z),\tau$.
Let us define the dimensionless measure of particle inertia as
\begin{equation}
\label{definition of I}
I(z)=\left(\frac{\tau}{ \tilde\tau(z)}\right)^{2/3}.
\end{equation}
Equivalently, $I(z)$ can be represented through the ratio of the particle mean free path and the scale of flow inhomogeneity.
For $\kappa$ given by (\ref{diffusivity}) the inertia degree is position independent, $I=(\mu\tau)^{1/3}$.

Our goal is to determine the long-term behaviour of inertial particles near the inelastic wall. 
Namely, we are interested in the direction of preferential particle migration.
In the limit $I\ll 1$ the approximation of local equilibrium is valid:  the statistics of particle velocity is described by the Maxwell distribution with  $\gamma\kappa(z)$ playing the role of local temperature.
Thus, the particles adjust velocity locally and do not sense the boundary conditions in the leading-order approximation.
The particle concentration in the real space $n(z,t)=\int_{-\infty}^{+\infty} dv\rho(z,v,t)$ on timescales $t\gg\tau$ obeys the  gradient transport equation $\partial_t n=-\partial_z j$, where the diffusion current $j$ is
\begin{equation}
j=-\kappa(z)\partial_z n-n\frac{d\kappa(z)}{dz}.
\end{equation}
The second term in the right hand side is proportional to the eddy diffusivity gradient and
represents  usual turbophoretic current in the direction of decreasing turbulence.
For quadratic $z$-dependence of effective diffusivity the fluxless steady-state $n(z)\propto 1/z^2$ is normalizable at $z\to\infty$ so that particles concentrate near minimum of turbulence.

For inertia degree $I$ of order $1$ and larger the particles are far from the equilibrium with local turbulence and gradient
transport equation for the particle real space concentration is invalid.
Recent theoretical analysis \citep{BFF} of quadratic model revealed a localization-delocalization transition upon the change of $I$ for inertial particles near ideally reflecting wall:  the direction of particle migration reverses as inertia degree is getting larger than some critical value $I_c\approx 0.827^{-1}$.
In the next section we present a theoretical approach allowing us to describe this transition in terms of particle trajectories.
In the Section \ref{sec:inelastic wall} the approach is extended to the case of inelastic wall.

\section{Particles near the perfectly reflecting wall}
\label{sec:minimum}

If the wall at $z=0$ is elastic (perfectly reflecting), the time dependence of the particle position for each realization of turbulence is given by
\begin{equation}
\label{particle coordinate}
z(t)=z(0)\exp\left(\principal\int_{0}^{t}\sigma(t')dt'\right),
\end{equation}
where $\sigma=v/z$.
The principal value ($\principal$) of the integral corresponds to the moments of particle-wall collisions $t_i$, for which variable $\sigma$ becomes singular. The representation (\ref{particle coordinate}) is valid for $t\ne t_i$.

We wish to determine the direction of particles migration.
The long-time evolution of $z$ is determined by the Lyapunov exponent
\begin{equation}
\label{Lyapunov exponent0}
\lambda=\lim\limits_{t\to +\infty}\frac{1}{t}\ln\frac{z(t)}{z(0)}=\lim\limits_{t\to +\infty}\frac{1}{t}\principal\int_{0}^{t}\sigma(t')dt'=\langle\sigma\rangle
\end{equation}
With the assumption of ergodicity, the time average converges to ensemble average, so that the mean value of $\sigma$ in (\ref{Lyapunov exponent0})  can be calculated as $\langle\sigma \rangle=\lim\limits_{t\to +\infty}\int_{-\infty}^{+\infty}\sigma P(\sigma,t)d\sigma$, where $P(\sigma,t)=\langle \delta(\sigma-\sigma(t))\rangle$. 
It follows from (\ref{FP1}) that probability distribution  $P$ satisfies the closed Fokker-Planck equation
 \begin{equation}\partial_tP=\partial_{\sigma}\left[U'P\right]+\mu\gamma^2\partial_{\sigma}^2 P\ ,\label{FPSigma}\end{equation}
with  $U(\sigma)= {\gamma\sigma^2}/{2}+ {\sigma^3}/{3}$.
Note, that dynamics of variable $\sigma$ decouples from that of $z$ and $v$ due to the quadratic $z$-dependence of diffusivity (\ref{diffusivity}).

The $\sigma$-space has the topology of a circle with $\sigma=-\infty$ glued to $\sigma=+\infty$ at $z=0$. Every reflection from $z=0$ corresponds to  $\sigma$ jumping from $=-\infty$ to $+\infty$.
That topology allows the flux of probability $-F$ going towards $-\infty$ and returning from  $+\infty$: $U'P+\mu\gamma^2P'=F$. The flux independent of $\sigma$ corresponds to the normalizable steady state,
\begin{equation}P(\sigma)={F\over\mu\gamma^2}e^{-U(\sigma)/\mu\gamma^2}\int_{-\infty}^{\sigma}\!\!\!
e^{U(\sigma')/\mu\gamma^2} d\sigma'\,,\label{Psigma}\end{equation}
which realizes the minimum of entropy production \citep{FT} and is the asymptotic state of any initial distribution \citep{Gaw2}.
The constant flux $F$ is the average frequency of particle-wall collisions and behaves as \citep{Gaw2}
\begin{eqnarray}
\label{collision rate}
F=\frac{ I}{\sqrt{\pi}\tau}\left[\int_{-\infty}^{+\infty}\frac{dx}{\sqrt{x}}\exp\left(-\frac{x^3}{12}+\frac{x}{4I^2}\right)\right]^{-1}=\frac{I}{\pi^2\tau}\left[\Ai\left(\frac{1}{4I^2}\right)^2+\Bi\left(\frac{1}{4I^2}\right)^2\right]^{-1}
\end{eqnarray}
where $\Ai$ and $\Bi$ designate the Airy function of the first
kind and second kind respectively.
This expression comes from normalization condition for  distribution (\ref{Psigma}), $\int_{-\infty}^{+\infty} P(\sigma)d\sigma=1$.

\begin{figure}
  \centerline{\includegraphics[scale=0.7]{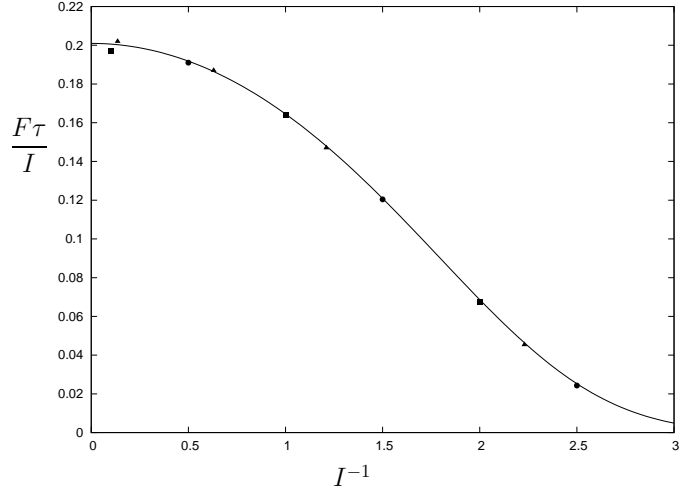}}
  \caption{The dependence of the collision frequency on the inertia degree $I$: analytical result (\ref{collision rate}) (line) and numerics for $\beta=1,$ ($\circ$), $0.5$ ($\blacksquare$) and $0.1$ ($\blacktriangle$).
  This plot convincingly demonstrates that collision rate does not depend on the coefficient of restitution $\beta$.
}
\label{fig:b}
\end{figure}

We can now calculate the Lyapunov exponent (\ref{Lyapunov exponent0}) by using steady-state
 PDF (\ref{Psigma})
\begin{eqnarray}
\label{Lyapunov exponent1}
\lambda(\beta=1,I)=\int_{-\infty}^{+\infty}\sigma P(\sigma)d\sigma= \frac{\sqrt{\pi}}{2}F\int_{0}^{+\infty}dx \frac{x-I^{-1}}{\sqrt{x}}\exp\left(-\frac{x^3}{12}+\frac{x}{4I^2}\right)
\end{eqnarray}
This integral is negative at $I<I_c\approx 0.827^{-1}$ and positive otherwise \citep{WM}.
 That sign change is a phase transition, called localization-delocalization transition for motion of particle near a wall (when $z$ is the distance from the wall) and path-coalescence transition for inter-particle dynamics (when $z$ is an inter-particle distance).
Note that even in the delocalized phase, despite going away exponentially,
$z(t)$ returns to zero with the time-independent mean rate (\ref{collision rate}).

The mean squares of the coordinate and the velocity, $\langle z^2(t)\rangle=\int z^2\rho(z,v,t)dzdv$ and $\langle v^2(t)\rangle=\int v^2\rho(z,v,t)dzdv$, obey the identical closed equations
 \begin{equation}
 \frac{d^3\langle z^2(t)\rangle}{dt^3}+3\gamma\frac{d^2\langle z^2(t)\rangle}{dt^2}+2\gamma^2\frac{d\langle z^2(t)\rangle}{dt}-4\gamma^3I^3\langle z^2(t)\rangle=0,
 \end{equation}
 \begin{equation}
 \frac{d^3\langle v^2(t)\rangle}{dt^3}+3\gamma\frac{d^2\langle v^2(t)\rangle}{dt^2}+2\gamma^2\frac{d\langle v^2(t)\rangle}{dt}-4\gamma^3I^3\langle v^2(t)\rangle=0,
 \end{equation}
which have solutions in the form $Ae^{\lambda_1 t}+Be^{\lambda_1 t}+Ce^{\lambda_3 t}$.
The amplitudes $A$, $B$ and $C$ can be expressed through the initial coordinate $z_0$, initial velocity $v_0$ and the eigenvalues $\lambda_1$, $\lambda_2$ and $\lambda_3$.
The latter parameters are completely determined by the inertia degree $I$ and provide the long-term exponential growth of both moments for any initial conditions.
 This growth may seem surprising in the case of localization; however, realizations with particles going away exist even in this case - those realization contribute negligibly  $\langle \ln z(t)\rangle$ but dominate $\langle z^2(t)\rangle$.

\section{The case of inelastic collisions}\label{sec:inelastic wall}

Assuming that at every collision with the wall the particle instantaneously loses some part of its wall-normal velocity we obtain  $v(t_i+\delta t)=-\beta_i v(t_i-\delta t)$ and $z(t_i+\delta t)=\beta_i z(t_i-\delta t)$, where $z(t_i)=0$ and $\delta t\to+0$. 
In general, $\beta_i$ should be viewed as an effective restitution coefficient related to energy dissipation due to collision-induced plastic deformations and particle-wall hydrodynamic interaction \citep{Legendre}.
The realistic models for these dissipation processes should involve $\beta_i$ which is a function of the impact velocity.
Note also that temporal or spatial irregularities at the surface of the wall may give rise to randomness of restitution coefficient. 
Remarkably, since $\sigma(t_i+\delta t)= -\sigma(t_i-\delta t)$, the variable $\sigma=v/z$ does not feel the boundary conditions at the wall.
This means that the probability density function $P(\sigma,t)$ relaxes to the same steady state (\ref{Psigma}) independently on details of particle-wall interactions. 

In this study we focus our attention on the idealized model with constant restitution coefficient $\beta$. 
Then, it follows from the relation $z(t_i+\delta t)=\beta z(t_i-\delta t)$ that the particle coordinate for  each particular realization of turbulence is given by
\begin{equation}
z(t)=z(0)\beta^{N(t)}\exp\left(\principal\int_{0}^{t}\sigma(t')dt'\right),
\end{equation}
where, as before, $N(t)$ is number of bounces up to the moment $t\ne t_i$.
%

The Lyapunov exponent can be then calculated as
\begin{equation}
\lambda(\beta, I)=\lim\limits_{t\to +\infty}\frac{1}{t}\ln\frac{z(t)}{z(0)}=\lim\limits_{t\to +\infty}\frac{1}{t}\principal\int_{0}^{t}\sigma(t')dt'
+\ln\beta\lim\limits_{t\to +\infty}\frac{N(t)}{t}.
\end{equation}

Since the statistics of $\sigma$ does not depend on $\beta$, the first term in the rhs is just the Lyapunov exponent (\ref{Lyapunov exponent1}) in the  problem with an ideally reflecting wall
\begin{equation}
\lim\limits_{t\to +\infty}\frac{1}{t}\principal\int_{0}^{t}\sigma(t')dt'=\lambda(1,I),
\end{equation}
while the second limit is the collision rate
 \begin{equation}
\lim\limits_{t\to +\infty}\frac{N(t)}{t}=F(I),
\end{equation}
which is given by (\ref{collision rate}).

\begin{figure}
 \centerline{\includegraphics[scale=0.7]{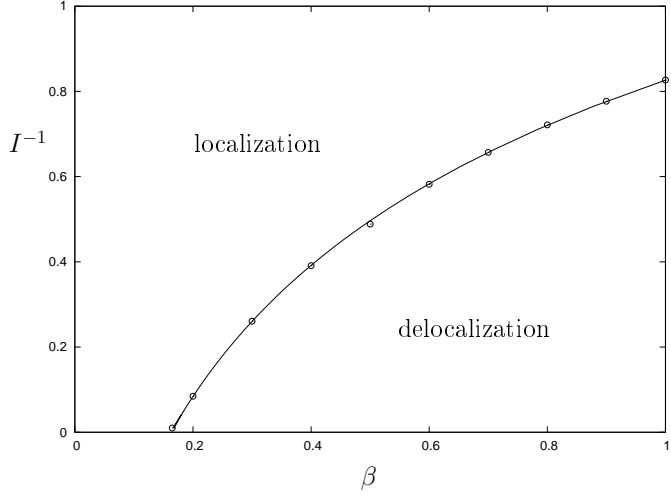}}
  \caption{The phase diagram of the localization-delocalization transition: theory (line) and numerical simulations ($\circ$).}
\label{fig:a}
\end{figure}

We then express the Lyapunov exponent for any inertia parameter $I$ and the restitution factor $\beta$:
\begin{equation}
 \lambda(\beta,I)=\lambda(1,I)+F(I)\ln\beta.
\end{equation}
Localization-delocalization phase transition corresponds to a zero Lyapunov exponent $\lambda(1,I)+ F(I)\ln\beta=0$
which determines the transition curve (see Figure \ref{fig:a})
\begin{eqnarray}
\ln \beta_c(I) =- \frac{\lambda(I)}{F(I)} =-\frac{\sqrt{\pi}}{2}\int_{0}^{+\infty}dx \frac{x-I^{-1}}{\sqrt{x}}\exp\left(-\frac{x^3}{12}+\frac{x}{4I^2}\right)\,,\nonumber
\end{eqnarray}

Calculating the integral for the infinite inertia parameter ($I\to\infty$) one obtains non-zero value
\begin{equation}
\beta_0=\exp\left(-\frac{\pi}{\sqrt{3}}\right)\ .\label{beta0}
\end{equation}
We thus conclude that the curve of the phase transition in the plane $\beta-I$ hits $I=\infty$ at the finite $\beta=\beta_0$.
There is no phase transition for less elastic walls where particles with any inertia localize.
By the same token, particles coalesce if their restitution coefficient is small enough.
This phenomenon can be called \emph{ inelastic collapse} in analogy to similar effect in driven granular matter \citep{CSB}.
Even though beautiful formula (\ref{beta0}) has been obtained within a short-correlated model, it perfectly corresponds to the direct numerical simulations presented in the next section for a synthetic velocity field with finite temporal correlations.

Interestingly enough, the present calculations can be extended to the case of non-deterministic particle-wall interaction.
Specifically, for a rough wall, one can treat the restitution coefficient of the wall-normal velocity $\beta$ as a random variable having some probability distribution function $p(\beta)$.
Then, the particle coordinate evolves as
\begin{equation}
\label{position random}
z(t)=z(0)\exp\left(\principal\int_{0}^{t}\sigma(t')dt'\right)\prod\limits_{i=1}^{N(t)}\beta_i,
\end{equation}
where $\beta_i$ is the random restitution coefficient associated with $i$th bounce.
We assume that random numbers $\beta_1,\beta_2,...,\beta_{N}$ are statistically independent.

From (\ref{position random}) one obtains the following expression for the Lyapunov exponent
\begin{equation}
\tilde\lambda=\lim\limits_{t\to +\infty}\frac{1}{t}\principal\int_{0}^{t}\sigma(t')dt'
+\lim\limits_{t\to +\infty}\frac{1}{t}\sum\limits_{i=1}^{N(t)} \ln\beta_i.
\end{equation}
As before, the first term in the rhs is the Lyapunov exponent (\ref{Lyapunov exponent1}) from the problem with elastic wall.
The second term can be calculated as
 \begin{equation}
\lim\limits_{t\to +\infty}\frac{1}{t} \sum\limits_{i=1}^{N(t)} \ln\beta_i=F(I)\lim\limits_{N\to +\infty}\frac{1}{N} \sum\limits_{i=1}^{N}\ln\beta_i=F(I) \int_{0}^{1} \ln \beta p(\beta) d\beta.
 \end{equation}
That finally gives
\begin{equation}
\tilde\lambda=\lambda(I) + F(I)\langle\ln \beta\rangle_p.
\end{equation}
If $\langle \ln \beta\rangle_p=\int_{0}^{1} \ln \beta p(\beta) d\beta>\ln \beta_0$, the particle is localized for any inertia degree $I$. 
Analysis of the localization properties for the case of a velocity-dependent restitution coefficient remains a task for the future.

Finally in this section, let us briefly discuss the effect of partial boundary absorption. 
We assume that on arriving at the wall the particle is absorbed with probability $\alpha$ and reflected with probability $1-\alpha$. 
The wall-normal  velocities of the particle just after and before reflection are related by $v\to-\beta v $, where restitution coefficient $\beta$ is a random variable or a given function of impact velocity.
We are interested in the survival probability $Q(t)$  which is the probability that a particle  has not yet been absorbed by wall after a time $t$.
The decay rate of $Q(t)$ is given by absorption probability $\alpha$ multiplied by the collision rate $F$: $dQ/dt=-\alpha F$.
At long timescales the probability distribution of variable $\sigma$ relaxes to steady state (\ref{Psigma}) providing the time-independent frequency of collisions $F$ given by (\ref{collision rate}).
This leads to the exponentially fast long-term decay  $Q(t)\propto \exp(-\alpha Ft)$.
Note that the survival probability does not depend on the restitution coefficient $\beta$ due to the aforementioned insensitivity of $\sigma$ to the boundary condition at the wall.

%

\section{Simulations}
We conducted numerical simulations of the motion of particles subjected to one-dimensional inhomogeneous random force.
In doing numerics we model the gradient of fluid velocity as a telegraph noise. That means that  we choose $u=z\xi(t)$, where the random process $u=z\xi(t)$ is a piece-wise random  constant during $\tau_c$, chosen from a Gaussian distribution with a zero mean and unit variance. Varying $\tau_c/\tau$ we are able to explore finite-correlated flows as well.
The equation of motion Eq.(\ref{Langeven}) is solved by the second-order Runge-Kutta algorithm with time steps much smaller than $\tau_c$. Wall reflection is modeled as follows: if a particle changes the sign of $z$ during $\tau_c$, its coordinate flipped to opposite while its velocity changes sign and is multiplied by the restitution coefficient $\beta$.
To provide the relevance of the numerical simulations to our analytical model, we have used the the correlation time of fluid velocity $\tau_c=0.002$, which is much smaller than other time parameters of the problem, $\tau_c\ll \tau, F^{-1},\lambda^{-1}$.
In this limit, the particle displacement during $\tau_c$ is small and the telegraph process reproduces the case of a white noise.

\begin{figure}
  \centerline{\includegraphics[scale=0.7]{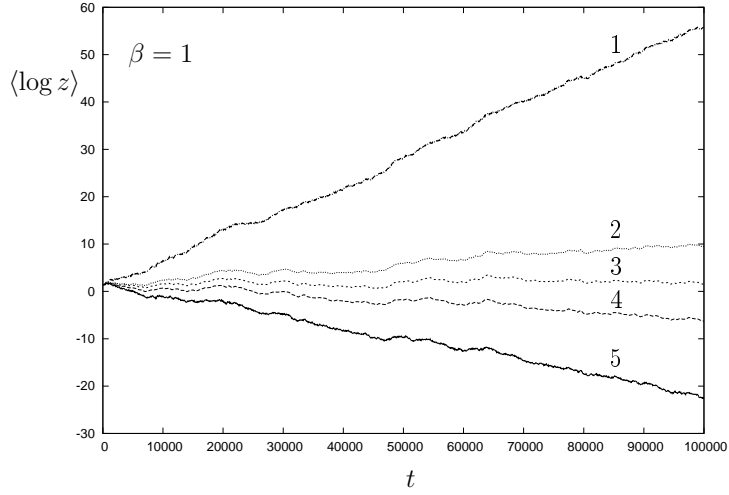}}
  \caption{The time dependence of the mean logarithm of the particle coordinate at fixed elasticity $\beta=1$ for different degrees of inertia:
  (\textit{1}) $I=1.25147$ , (\textit{2}) $1.21545$,  (\textit{3}) $1.20939$, (\textit{4}) $ 1.20332$ and (\textit{5}) $ 1.19144$.}
\label{fig:c}
\end{figure}

\begin{figure}
  \centerline{\includegraphics[scale=0.7]{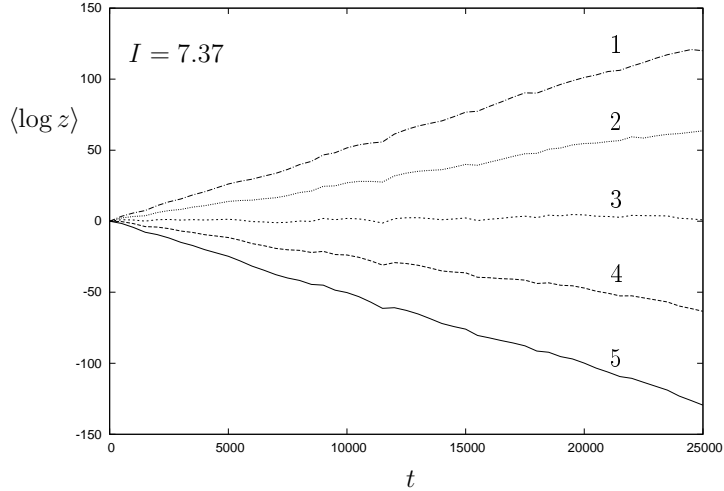}}
  \caption{The time dependence of the mean logarithm of the particle coordinate at fixed inertia degree $I=7.37$ for different restitution coefficients:
  (\textit{1}) $\beta=0.21$ , (\textit{2}) $0.205$,  (\textit{3}) $0.20$, (\textit{4}) $ 0.195$ and (\textit{5}) $0.19$.}
\label{fig:d}
\end{figure}

We compute Lyapunov exponents to determine the direction of the particle drift
\begin{equation}
\lambda=\lim_{t\to \infty}\frac1t\left\langle \ln \frac{z(t)}{z(0)}\right\rangle,
\end{equation}
where averaging is over many (typically 1000) realization of the random process.
The Lyapunov exponent is known to be self-averaging quantity, that is having the same value in any given long realization as after averaging over many realizations.
The localization-delocalization phase transition occurs when the Lyapunov exponent changes its sign.

The figure \ref{fig:b} indicates that, as expected, the collision frequency $F$ does not depend on $\beta$ and is completely determined by the inertia degree $I$ and relaxation time $\tau$.
Figures \ref{fig:c} and \ref{fig:d} show that the mean logarithm of particle coordinate is a linear function of time.
The slope changes its sign upon the change of inertia degree or elasticity.
The phase diagram  of localization-delocalization transition demonstrates a very good agreement of numerics and analytic theory, see Fig. \ref{fig:a}.

\section{Phase transition upon the change of the Stokes number}\label{sec:Stokes}

As stated above, there are three time parameters for an inertial particle in a random flow: the velocity correlation time $\tau_c$ of the fluid, the particle relaxation time $\tau$, the time needed by the particle to feel the flow inhomogeneity $\tilde\tau$.
In the previous sections the transition upon the change of inertia parameter $I=(\tau/\tilde\tau)^{2/3}$ was demonstrated in the limit $\tau_c\ll \tau,\tilde \tau$.
Here we consider $\tau_c,\tau\ll \tilde \tau$ and discuss the phase transition upon the change of the Stokes number $\St=\tau/\tau_c$.
The particles concentration on long timescales ($t\gg \tau,\tau_c$) obeys the equation  \citep{BFF}
 \begin{equation}
\partial_t n=\frac{\St}{1+\St}\partial_z^2[\kappa(z)n]+\frac{1}{1+\St}\partial_z[\kappa(z)\partial_z n]\ .\label{St1}
\end{equation}
For quadratic diffusivity profile $\kappa(z)=\mu z^2$ this leads to the following expression for the average logarithm of particle coordinate $\langle \ln z(t)\rangle=\int  n(z,t)\ln z\, dz$
 \begin{equation}
\partial_t \langle \ln z(t)\rangle=\lambda=\mu \frac{1-\St}{1+\St}
\end{equation}
At the critical Stokes number $\St_c=1$ the Lyapunov exponent vanishes and localization-delocalization transition occurs. The reason of this transition is that at low particle inertia ($\St< \St_c$) the turbophoresis can not compensate the effect of turbulent diffusion which tends to spread the particles throughout the fluid.
On the plane $x=I,y=1/\St$ we expect the phase-transition curve to start at $x=I_c,y=0$ and monotonically go down to $x=0, y=1$. This is supported by the numerical data obtained with increasing ration of the flow correlation time to the Stokes time. Taking $\beta=1$, we find $I_c=1.209$ for $\St=500$,  $I_c=1.204$ for $\St=50$ and  $I_c=1.137$ for $\St=5$. It is important to keep in mind that  an account of fluid incompressibility (and multi-dimensionality) was crucial in deriving (\ref{St1}) by \cite{BFF}.
Therefore,  the motion of particles with $\tau\lesssim \tau_c$ cannot be modeled numerically with a one-dimensional fluid flow.

\section{Discussion}

We have studied analytically and numerically the relative motion of inertial particles in homogeneous random flow or, equivalently, the particle motion in inhomogeneous wall-bounded flow.
The main part of our analysis was focused on the case of a short-correlated flow with
quadratic dependence of the effective diffusivity on the distance, which corresponds to the inter-particle distance in the viscous interval of turbulence or to the motion of a single particle near a minimum without a wall.
In this model, the direction of particle migration is determined by two dimensionless parameters: the inertia degree $I$ and the restitution coefficient of particle-particle or particle-wall collisions $\beta$.
We have derived analytically the phase diagram for the localization-delocalization transition in the plane $I-\beta$.
Remarkably,  there is a critical value of the coefficient $\beta$ below which particles localize for any value of inertia $I$.
The results are confirmed by the direct numerical simulations of particle motion in a synthetic random flow.

Our results do not apply directly to the near-wall region of developed turbulence, where the turbulent diffusivity behaves as $z^4$ inside the viscous sub-layer and as $z$ in the logarithmic boundary layer.
An interesting question is whether the results provided by exactly solvable quadratic model can be generalized to these cases.
Recent analysis \citep{LB} of inertial particles in the viscous sub-layer of wall-bounded turbulence ($\kappa\propto z^4$) revealed the localization-delocalization transition at the same $\beta=\beta_c$.
 For  $\kappa\propto z^m$ we thus have the same threshold for localization at $m=2$ and $m=4$. Moreover, the critical value (\ref{beta0}) is the same for inelastic collapse of trajectories under the action of a spatially uniform random force, $\kappa(z)=$const, as long as it is zero upon the contact, $\kappa(0)=0$ \citep{CSB}.
To address the localization properties of inertial particles for any $m\ge 0$ one needs to consider the boundary value problem (\ref{FP1}-\ref{boundary condition}), which will be the subject of future work.

It is important to stress that our one-dimensional consideration with a short-correlated flow is a crude approximation of reality, particularly for the problem of two inertial particles.  Further numerical and experimental work is needed to establish whether there is some analogue of the
collapse transition leading to clustering of particles in real flows. This seems likely as one would expect that inelastic inter-particle collisions decrease the Lyapunov exponent associated with
the evolution of distance between two inertial particles below the viscous scale of turbulence.

The work was supported by the Minerva Foundation with funding from the German Ministry for Education and Research. The work of GF and SB (analytic theory and writing the paper) was supported by the RScF grant 14-22-00259.

\bibliographystyle{jfm}

 {}


\end{document}